\documentclass[preprint,aps,prc,showpacs]{revtex4}  
\usepackage{times}
\usepackage[dvips]{graphicx}
\newcommand{\be}{\begin{eqnarray}}
\newcommand{\ee}{\end{eqnarray}}

\begin{document}
\title{ON THE CHEMICAL FREEZE-OUT CRITERIA IN A HOT AND DENSE FIREBALL}
\vskip 0.4in
\author{M. Mishra} 
\email{madhukar.12@gmail.com}
\affiliation{Department of Physics, Banaras Hindu University, Varanasi 221005, India}
\author{C. P. Singh}
\email{cpsingh_bhu@yahoo.co.in}
\affiliation{Department of Physics, Banaras Hindu University, Varanasi 221005, India}
\vspace{12pt}

\begin{abstract}

  Intensive investigations of freeze-out criteria in a hot and dense fireball provide important information regarding particle emission from the fireball. A systematic comparison of these proposals is presented here in the framework of a thermodynamically consistent excluded volume model which has been found to describe the properties of hadron gas (HG) quite well. We find that the impact of excluded volume correction is considerably large and the average energy per hadron is $0.9$ GeV, $n_b+n_{\bar b}$ stays nearly constant at $0.12$/fm$^3$ and the normalized entropy density $s/T^3\approx 6$ in this model. Moreover, these values are independent of the beam or center-of-mass energy and also of the target and beam nuclei. In ideal HG model these quantities show substantial energy dependence. Further we have compared the predictions of various excluded volume models in the precise determination of these criteria and we find that the thermodynamically consistent excluded volume models give the best results. In addition, we find another important criterion that entropy per hadron has a constant value at $6$ in our model. We hope that these findings will throw considerable light on the expansion dynamics and the bulk thermodynamic properties of the fireball before chemical freeze-out.        
\end{abstract}


\maketitle
\section{Introduction}
Quantum chromodynamics (QCD) predicts a phase transition from a hot and dense hadron gas (HG) to a deconfined phase of quark gluon plasma (QGP) at very large temperature and / or baryon density. However, in spite of a considerable work in the past three decades, quantitative predictions for many aspects of this phase transition are still missing~\cite{tann1}. One important feature of paramount interest in the search of QGP is that it requires the use of many overlapping areas in physics like particle and nuclear physics, relativistic quantum mechanics and statistical mechanics, astrophysics and cosmology etc. The major problem one faces in the detection of QGP is how to relate the thermodynamical properties (temperature, energy density, entropy etc.) of the fireball produced in the relativistic heavy-ion collisions to the measurable properties observed in the laboratory. One finds that the matter created at RHIC appears to be more like a perfect liquid rather than a gas of free quarks and gluons. Here one gets a surprising result that the experimental result on the hadronic multiplicities can suitably be described by a thermal hadron gas model which describes the hot and dense, chemically equilibrated hadron gas emerging from the fireball but this picture, however, does not throw any light on the QGP formation before hadronization. Various authors have used models to explain hadron multiplicities and their ratios in terms of the chemical freeze-out parameters such as temperature $T$ and baryon chemical potential $\mu_B$ etc. of the fireball. The purpose of this paper is to search certain uniform conditions satisfied by the thermodynamic properties of the fireball at the chemical freeze-out. This investigation will provide us an insight into the particle emission properties of the fireball.

    Recently many papers have appeared~\cite{jcley2,jcley3,jcley4,pbra5,vmag6,jcley7,ataw8} which indicate following empirical conditions to be valid on the entire freeze-out surface of the fireball produced in the ultra relativistic heavy ion collisions at chemical freeze-out:\\
(a) a fixed value for the energy per hadron $E/N$ (i.e., ratio of energy density and number density $\frac{\varepsilon}{n})\approx 1.08$ GeV~\cite{jcley2,jcley3,jcley4};\\
(b) a fixed value for the sum of baryon and anti baryon densities $n_b+n_{\bar b}\approx 0.12$/fm$^3$~\cite{pbra5};\\
(c) a fixed value for the normalized entropy density $s/T^3\approx 7$~\cite{vmag6,jcley7,ataw8}.\\
It is certainly surprising to notice that each of the above conditions separately gives a satisfactory description of the chemical freeze-out points in the hot and dense fireball in an ideal hadron gas (HG) picture of the statistical thermodynamics. It is not yet clear what kind of relevant information regarding the physical properties of the fireball and its expansion dynamics we can achieve by these conditions. However,  the ideal HG model is unsuitable to describe the properties of the fireball at extremely large temperatures and/or densities~\cite{mmis9,cpsin10,sud11}. Moreover, Cleymans et al.,~\cite{jcley3} have shown that incorporation of excluded volume correction leads to somewhat wild as well as disastrous effects on these conditions.          
We have recently proposed an excluded volume model~\cite{mmis9} which gives a thermodynamically consistent description of various thermodynamical properties of the fireball. The purpose of this paper is to investigate the nature of the above empirical conditions in our excluded volume model~\cite{mmis9}. After re-examination we find that these conditions involve modified constant values and show a remarkable independence so far as the collision energy, collision-geometry and the structure of the beam-target nuclei are concerned.
 
   The thermal statistical models have provided the most suitable description of all the experimental data available on the particle multiplicities as well as particle ratios from the lowest SIS to the highest RHIC energies. In these models, we assume that a hot and dense fireball is first produced in the heavy ion collisions and then it subsequently expands, cools and disintegrates into various hadrons, leptons and photons etc. At the chemical freeze-out point, the matter-composition in the fireball is frozen and then the fireball suffers a thermal freeze-out at which particles undergo their last elastic collisions. It is obvious that the thermal freeze-out occurs at a temperature lower than the chemical freeze-out, since the expansion of the fireball continues further. In a thermal model description, the observed multiplicities of the final hadrons emerging from the fireball are found to be consistent with the assumption of the chemical equilibrium existing in the fireball. In their description of the data at chemical equilibrium, the thermal models always involve two parameters only i.e., the temperature $T$ and baryon chemical potential $\mu_B$ at a given center-of-mass energy of the heavy-ion collisions. The partition function used in these calculations has also found support from results obtained in the lattice gauge theory~\cite{cra12,fka13} for the hadronic phase. However, in a simple treatment of HG, all the baryons as well as mesons are considered as non-interacting and point-like objects which give rise to an undesirable feature that at very high baryon density or large baryon chemical potential, the hadronic phase reappears as a stable configuration in the Gibbs construction of phase equilibrium between the hadron gas (HG) and the quark gluon plasma (QGP) phase~\cite{mmis9}. Moreover, this assumption implies that at very large $T$ or $\mu_B$, an arbitrarily large number of hadrons can be produced in a given volume $V$, and hence the energy density and pressure can rise to an infinitely large value. In order to remove these difficulties, excluded volume models have been invented in which each baryon is provided with a finite, proper hard-core volume. We have recently proposed~\cite{mmis9} a thermodynamically consistent excluded volume model and we find that the equation of state (EOS) of HG is described in the most reasonable fashion. It has many additional thermodynamically significant features, e.g., it tells us explicitly the difference between consistent and inconsistent models. It is found to be work even for the cases of extremely small as well as large temperatures and densities where most of the other approaches fail. It is also encouraging to see that its predictions on various quantities are completely in agreement with those given by a simulation model~\cite{nsas14} involving microscopic details of the collision geometry and the multiple scattering algorithm. We have also found that our model predicts meson and baryon densities in excellent agreement~\cite{mmis15} with the data obtained from the analysis of the pion interferometry (HBT)~\cite{dad16}. In this paper we have evaluated the fireball volume at chemical freeze-out and the extremely large value of the fireball volume indicates the existence of a mixed phase between the QGP to HG phase transition~\cite{mmis15}. These results make it amply clear that our excluded volume model can provide a reasonable description of the EOS of HG produced in the heavy-ion collisions from the lowest SIS to the highest RHIC energy. 
\section{formulation}
  In our model the fraction of occupied volume $R$ is given as\cite{mmis9}:
\begin{eqnarray}
R=\frac{\sum_i\,X_i}{1+\sum\,X_i}-\frac{\sum_i\,X_i^2}{(1+\sum_i\,X_i)^3}+
2\frac{\sum_i\,X_i^3}{(1+\sum_i\,X_i)^4}\nonumber \\
-3\frac{\sum_i\,X_i\,\lambda_i\sum_i\,
X_i^2\,I_i\,V_i}{(1+\sum_i\,X_i)^5}.
\end{eqnarray}
 
Using (1), one can calculate in excluded volume model, baryonic pressure as 
\begin{equation}
P_B^{ex}=(1-R)\sum P_i^0,
\end{equation}
where $P_i^0$ is the pressure in the point-like baryon approximation, $X_i=I_i\,\lambda_i\,V_i$ with $\lambda_i$ as the fugacity and $V_i$ the eigen volume of the $i^{th}$ baryonic species, and
\begin{equation}
I_i=\frac{g_i}{6\pi^2\,T}\,\int_0^{\infty}\,\frac{k^4\,dk}{\sqrt{k^2+m_i^2}}\,exp(-\sqrt{k^2+m_i^2}/T),
\end{equation}
here $g_i$ is the degeneracy factor of the $i^{th}$ particle. From (2), one can calculate baryon density, entropy density and energy density etc.
The total pressure of the HG is
\begin{equation}
P=P_B^{ex}+P_M^0,
\end{equation}
where $P_M^0$ is the pressure given by mesons in the point like approximation. In our model, we assign a hard core volume to baryons only. 

   We make use of the following parameterization to express $T$ and $\mu_B$ in terms of the center-of-mass energy $\sqrt{s_{NN}}$:
\begin{eqnarray}
\mu_B=\frac{a}{1+b\,\sqrt{s_{NN}}},\\
T=c-d\,e^{-f\,\sqrt{s_{NN}}},
\end{eqnarray}
where $a=1.308\pm0.028$ GeV, $b=0.273\pm0.008$ GeV$^{-1}$ and $c=172.3\pm 2.8$ MeV, $d=149.5\pm 5.7$ MeV and $f=0.20\pm0.03$ GeV ~\cite{jcley4,oan17,alx18,jcley19}. Here one must point out that the $T$, $\mu_B$ values as obtained in our model by fitting the experimental data on various hadron multiplicities do not differ much from those obtained in ideal hadron gas model. Thus we assume that the above parametrizations provide a valid description in our model also. However, one must add here that at very small values of $\sqrt{s_{NN}}$ ($<10$ GeV) the variations in $T$ and $\mu_B$ values are appreciable and are thus model dependent. Additionally we have used equal hard-core volume for all the baryons as  $V=4/3\,\pi\,r^3$ with $r=0.8$ fm~\cite{mmis9}. We do not give mesons any hard-core volume because bosons can easily overlap on each other and usually result in boson condensation. We have used all baryons, mesons and their resonances with a cut-off mass at $2$ GeV/c$^2$ in our calculation. We have also employed the strangeness conservation condition in the HG.
\begin{figure*}[tbp]
\begin{center} 
\mbox{\includegraphics[scale=0.8]{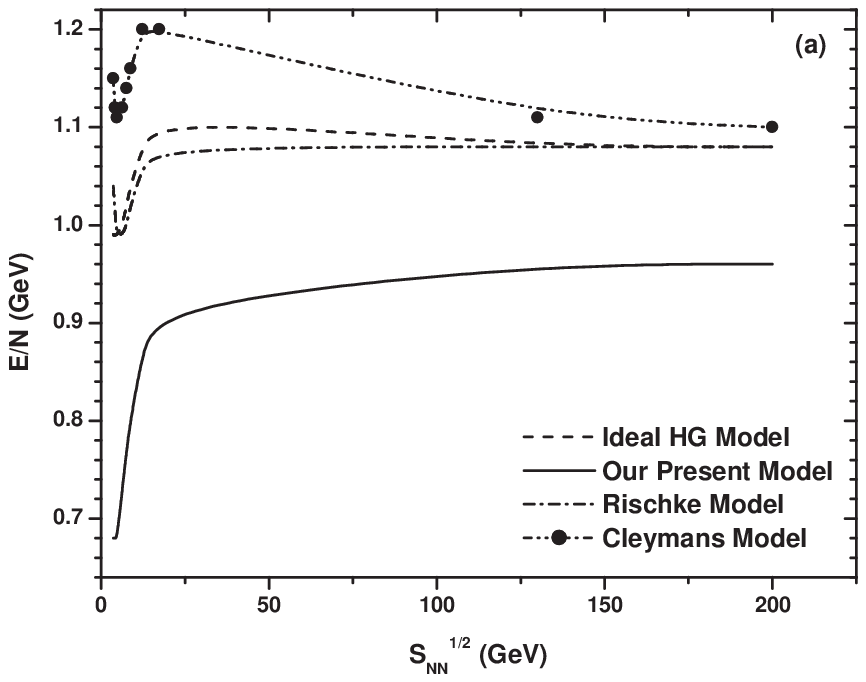} 
\includegraphics[scale=0.8]{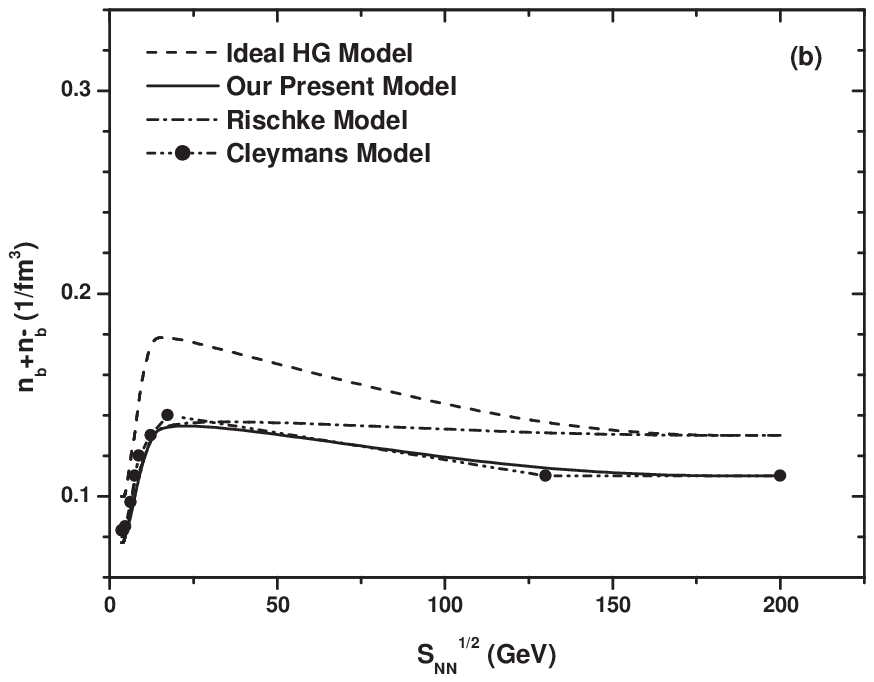}}
\mbox{\includegraphics[scale=0.8]{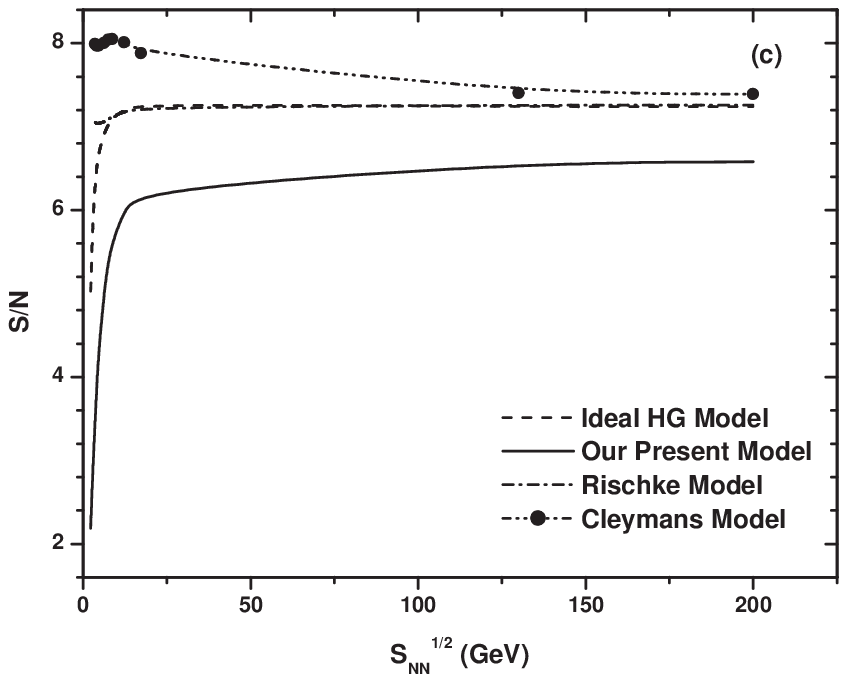} 
\includegraphics[scale=0.8]{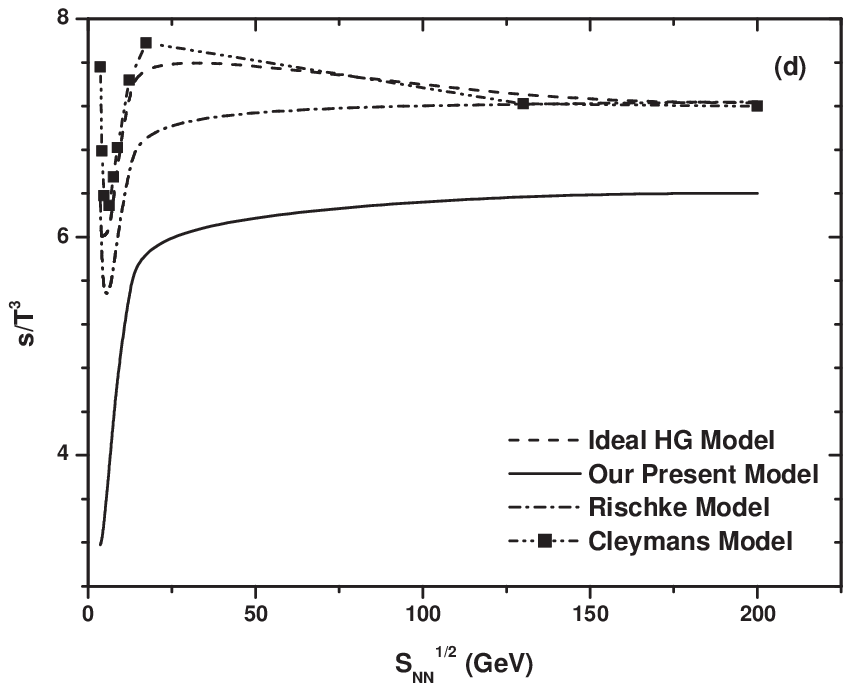}}
\caption{(a) Variation of $E/N$ with $\sqrt{s_{NN}}$. (b) Variation of $n_b+n_{\bar b}$ with $\sqrt{s_{NN}}$. (c) Variation of $S/N$ with $\sqrt{s_{NN}}$. (d) Variation of $s/T^3$ with $\sqrt{s_{NN}}$. Ideal HG calculation is shown by dotted curve and calculation in our model is given by solid line. Rischke et al., model and Cleymans-Suhonen model predictions are depicted by dash dotted curve and dash dot dot curve with solid circles, respectively.}
\end{center}
\end{figure*}
\section{Results and Discussions}
  The results of our calculation are shown in Fig. 1. Fig. 1(a) shows the variation of energy carried by each hadron i.e., $E/N$ with respect to  $\sqrt{s_{NN}}$ at chemical freeze-out point of the fireball. The ratio  $E/N$ shows indeed a constant value of $0.9$ GeV in our calculation and it shows a remarkable energy independence. In comparison, the curve in the ideal gas gives $E/N=1.08$ and involves a small variation with $\sqrt{s_{NN}}$. We have also shown the results in the Cleymans-Suhonen model~\cite{jcley20} as well as in the Rischke model~\cite{drisch21} where the former is a thermodynamically inconsistent model but the latter provides a thermodynamically consistent description of the fireball. We find that Rischke model also gives a constant value of $E/N$ as $1.08$ but Cleymans model calculation does not involve such energy independence for $E/N$.   

   In Fig. 1(b), we show the variation of $n_b+n_{\bar b}$ with $\sqrt{s_{NN}}$. In our model, this quantity has approximately a value of $0.12$/fm$^3$ but involves a small energy dependence. However, in ideal gas model this quantity involves a much larger variation with respect to center-of-mass energy and $n_b+n_{\bar b}$ is always larger than $0.12$/fm$^3$. In Cleymans-Suhonen model, the results almost coincide with our model. In contrast $n_b+n_{\bar b}=0.12$/fm$^3$ in Rischke model and it shows a complete energy-independent behaviour. P. Braun-Munzinger et al.,~\cite{pbrau22} used the excluded volume model of Rischke et al.,~\cite{drisch21} and found similar result for $n_b+n_{\bar b}$. However, they have used different eigen volumes for baryons and mesons. Here we would like to point out that we have plotted total baryon density with respect to collision energy in order to show that density remains constant and thus it gives a better method to judge the validity of freeze-out criterion. Moreover, we have compared the predictions of all types of models in order to show the applicability of excluded volume corrections accounted in the thermodynamically consistent manner.      

   In Fig. 1(c), we have demonstrated the remarkable energy independence shown by the quantity entropy per particle i.e., $S/N$ in our calculation. It has a constant value $S/N=6$. However, $S/N=7$ in the ideal HG model and is again almost independent of collision energy. 
The obtained results in Rischke model again shows a constant value for $S/N=7$ and it remains independent of collision energy as well as participating nuclei. Ideal HG model results overlap completely on the Rischke model curve. However, thermodynamically inconsistent model of Cleymans and Suhonen gives $S/N> 7$ and shows much energy dependence.

   Fig. 1(d) gives the center-of-mass energy variation of the normalized entropy density $s/T^3$ which reveals surprisingly similar variation as given by $S/N$ in Fig. 1(c). In our model, we get $s/T^3=6$ and it is remarkably energy-independent. Similarly in Rischke model the value of $s/T^3=7$ and remains constant. However, ideal HG model involves energy-dependence and the value is always larger than $7$. Moreover, Cleymans-Suhonen model result almost coincides with the ideal gas behaviour. The results shown in Fig. 1 demonstrate that all these freeze-out criteria show a large fluctuation in their values at very small $\sqrt{s_{NN}}$ and this indicates that the grand canonical partition function approach is probably insuitable for describing the freeze-out criteria in these models. 

   In this paper, we thus attempted to examine the freeze-out criteria in heavy-ion collisions. We have shown results for these criteria in various excluded volume models used in the literature. We find that all these criteria present a unified description in all thermodynamically consistent excluded volume models since the values of $E/N$, $n_b+n_{\bar b}$ and $s/T^3$ show a remarkable energy independent behaviour. We have also invented that a constant value for $S/N$ demonstrate another good freeze-out criterion in these models. Actual values differ in different models and are thus model dependent. For example, $S/N$ in our model has a value $6$. However, $S/N$ in the model of Rischke et al., has a value $7$. But all these criteria do not show such kind of energy independence in either ideal HG model or in Cleymans Suhonen excluded volume model which lacks thermodynamical consistency. The empirical conditions as shown in fig. 1 show a much better energy independence in the excluded volume model. We find that these results are completely opposite to the claim made by Cleymans et al.~\cite{jcley3}. They have observed that the incorporation of excluded volume correction leads to somewhat wild behaviour demonstrated by these quantities. In ideal HG consisting of massless pions only, $s_{\pi}=\frac{\varepsilon_{\pi}+P_{\pi}}{T}$ and $n_{\pi}=\frac{P_{\pi}}{T}$. So $\frac{s_{\pi}}{n_{\pi}}\approx 4$. We also know that baryons carry more (almost twice) entropy in comparison to mesons. Thus the average value of $s/n\approx 6$ for the HG as obtained in our model is quite consistent to our expectations.    
   
   In conclusion, we have presented a systematic comparison of all the freeze-out criteria proposed by various authors and we find that the  freeze-out criteria depend crucially on the excluded volume correction used in the ideal HG model. However, reanalysis of these conditions in our model as well as in Rischke model shows much better energy independence and all conditions are also independent of beam and target nuclei. Our finding lends support to the crucial assumption of chemical equilibrium achieved in the thermal model calculations. We hope that these conditions will help us in unveiling the physics of the fireball expansion in heavy-ion collisions. These findings~\cite{mmis9,cpsin10,sud11,cra12,fka13,nsas14,mmis15} give us confidence in the equation of state (EOS) developed by us for HG, as it gives a realistic picture of the hot, dense fireball produced in heavy-ion collisions. The conditions obtained here are significant in the evolution of the fireball and warrant further study on their physics implications. 
  
\begin{acknowledgements}
    M. Mishra is grateful to the Council of Scientific and Industrial Research (CSIR), New Delhi for his financial assistance. We would also like to thank Dr. V. J. Menon for many stimulating discussions and helpful suggestions.
\end{acknowledgements}

\end{document}